# Nanoplasmonic Optical Fiber Sensing of SARS-CoV-2 Nucleocapsid Protein Using an Aptamer-DNA Tetrahedron Interface

Xu Pin[1#], Cui Jingyu[2#], Cheng Zhi[1], Simon Chi-Chin Shiu[2], Cui Jingxian[1], Li Yujian[1], Liu Yifan[1], Wang Lin[2, 3], Ryan Ho Ping Siu[2], Julian A. Tanner[2, 3, 4*], and Yu Changyuan[1*]

[1]Department of Electrical and Electronic Engineering, The Hong Kong Polytechnic University, Hong Kong SAR, P.R. China

[2]School of Biomedical Sciences, Li Ka Shing Faculty of Medicine, The University of Hong Kong, Hong Kong SAR, P.R. China

[3]Advanced Biomedical Instrumentation Centre, Hong Kong Science Park, Shatin, New Territories, Hong Kong SAR, P.R. China

[4]Materials Innovation Institute for Life Sciences and Energy (MILES), HKU-SIRI, Shenzhen, P.R. China

*Correspondence Author (E-mail: jatanner@hku.hk, changyuan.yu@polyu.edu.hk)

#Xu Pin and Cui Jingyu contributed equally to this paper.

**Abstract** Optical fiber sensing carries a number of potential advantages for diagnostics and biomarker detection and monitoring, yet particular challenges persist in linking molecular recognition events to a change in the refractive index. DNA aptamers carry particular advantages as functional surface molecules on optical fibers to tailor detection of specific biomolecules, yet challenges persist around sensitivity and specificity. Diagnosis of COVID-19 through detection of nucleocapsid protein (N protein) of SARS-CoV-2 provides a classic diagnostic challenge where optical fiber-based sensing could complement and improve on typical detection methods such as RT-PCR and rapid antigen testing. In this study, a plasmonic gold-coated tilted fiber Bragg grating (TFBG)-based optical biosensing platform was developed for ultrasensitive detection of SARS-CoV-2 N protein. By functionalizing the optical fiber surface with aptamers for the molecular recognition of N protein, changes in refractive index measured biomolecular binding, thereby achieving real-time, label-free detection. Additionally, integrating DNA nanostructures such as the DNA tetrahedron with aptamers significantly enhanced detection sensitivity, increasing signal intensity ~2.5 times compared to aptamers alone. This study provides new insights into the development of high-performance optical fiber sensing platforms which integrate DNA nanostructure interfaces to facilitate biomarker recognition and sensing.

## 1. Introduction

Optical biosensors, leveraging the intrinsic interactions between light and matter alongside advancements in optoelectronic devices, hold promise for quantitative detection of disease biomarkers and non-destructive analysis. Such sensors integrate various optical techniques, including optical interferometry,[1] surface-enhanced Raman spectroscopy (SERS),[2,3] fluorescence,[4] surface plasmon resonance (SPR),[5,6] and colorimetric assay[7], providing diverse diagnostic solutions for biological detection. Among these techniques, SPR biosensors have garnered significant attention due to their high sensitivity, label-free detection, and real-time monitoring capabilities, making them a focal point of research.[8] Compared to conventional SPR systems, optical fiber-based SPR sensors enhance biosensing performance by compressing surface plasmon polariton (SPP) modes to the microscale, thereby achieving efficient energy interaction between the modal field and SPPs. To optimize detection performance, researchers have explored various fiber configurations, such as unclad,[9] side-polished,[10] tapered,[11] U shaped[12] and grating optical fiber, to precisely sense changes in the surrounding refractive index (SRI). Among these designs, plasmonic tilted fiber Bragg grating (TFBG)-based configurations stand out with the unique modal coupling mechanism, offering self-referencing capabilities and high resolution with the cladding modes and core mode of extremely narrow width. The dense spectral comb of cladding modes act as a vernier caliper during SPR excitation[13] creating a high-density comb of narrowband spectral resonances with a Q-factor of $10^4$,[14] and providing outstanding resolution ($10^{-6}$ to $10^{-8}$ RIU) for biological detection.[15] These features demonstrate exceptional potential in biological detection, enabling precise analysis of complex biomolecules and opening new possibilities for advanced biosensing applications.

One significant challenge for linking molecular recognition events to a change in the refractive index, is the development of an effective interface for biomarker recognition and sensing. Traditionally, optical fiber-based biosensors are functionalized with antibodies and rely on immunoassays to capture biomarkers. However, antibodies are often expensive, sensitive to temperature fluctuations, and prone to conformational changes upon binding to metallic surfaces.[16,17] To address these limitations, aptamers, often referred to as "chemical antibodies", have emerged as a promising alternative[18]. Aptamers were first introduced in 1990 through groundbreaking studies by Ellington and Szostak[19], Tuerk and Gold[20]. Employing in vitro evolution methodologies, they demonstrated the exceptional target-binding specificity of these nucleic acids. Owing to their distinctive properties, including high affinity, programmability, chemical stability, and facile modification-aptamers have emerged as pivotal tools in biosensing applications, offering a potential alternative to antibodies. However, despite these merits, aptamer-functionalized optical fiber sensors face significant challenges in practical deployment. A principal challenge arises from the heterogeneous distribution of aptamers across the fiber surface and constrained target molecule accessibility, which collectively hinder optimal aptamer alignment and exposure, thereby limiting signal amplification and reducing sensor sensitivity.

To address these limitations, DNA nanostructure technology has been extensively explored as an innovative approach to enhance the performance of aptamers. By leveraging the predictable base-pairing interactions of DNA strands, DNA nanostructures enable the construction of highly ordered and programmable architectures[21]. Among various DNA nanostructures, DNA tetrahedra have been widely utilized due to their simple and efficient synthesis, ease of design, and structural stability[22]. They form a stable three-dimensional rigid structure through the annealing and self-assembly of four oligonucleotide strands, which can integrate aptamers and provide an ideal scaffold, thereby significantly improving their performance. Studies have demonstrated that incorporating aptamers into DNA nanostructures allows for the precise control of their spatial arrangement and orientation, minimizing steric hindrance and maximizing accessibility to target molecules[23]. This optimization not only enhances the binding affinity and specificity of aptamers but also significantly improves the sensitivity of biosensing platforms.

Coronaviruses have emerged as a significant global health threat, triggering multiple outbreaks, with COVID-19 alone resulting in nearly 7.1 million deaths worldwide. Accurate identification of viral targets and implementation of effective diagnostic strategies are crucial for controlling viral transmission. The nucleocapsid protein (N protein) encoded by the coronavirus genome is considered an ideal diagnostic target due to its high abundance, widespread expression, and high conservation across viral variants[24]. Current methods for detecting N protein include RT-PCR, serological testing, and rapid antigen testing, but each has limitations: RT-PCR, while highly sensitive, relies on laboratory infrastructure; serological testing is less effective for early detection; and rapid antigen testing may lack sensitivity at low viral loads.[25-27] Therefore, developing a highly sensitive, specific, and point of care method for N protein detection is of great importance for controlling coronavirus spread and improving clinical diagnostics. Given these challenges, there is a growing need for complementary diagnostic technologies.

In this study, we developed an optical sensor integrating DNA nanostructures and aptamers, utilizing a real-time, in situ, label-free biofunctionalized 18° TFBG-SPR platform to detect the N protein. Compared to sensors based solely on aptamers, this sensor exhibits significantly enhanced sensitivity, improving N protein detection performance through structural optimization and signal amplification while preserving the aptamers' specific binding affinity. The findings robustly demonstrate the potential of DNA nanostructure-aptamer composites in advancing next-generation biosensors, offering promising prospects for applications in medical diagnostics and environmental monitoring.

## 2. Materials and methods

### 2.1. Materials and setup

Phosphate Buffer Saline (PBS), Tris(2-chloroethyl) phosphate (TCEP), and 6-mercapto-1-hexanol (MCH) solution were purchased from Beijing Solarbio Science & Technology Co., Ltd. (Beijing, China). Thiolated DNA strands were ordered from Integrated DNA Technologies (IDT). The nucleocapsid (N) protein was obtained from Sinobiological Inc. (Beijing, China). The Sodium Dodecyl Sulfate (SDS) and artificial saliva were obtained from Sigma-Aldrich (St. Louis, MO, USA).

### 2.2. Gold-coated tilted fiber Bragg grating fabrication

The fabrication of TFBGs was mainly performed with two steps, photosensitivity enhancement and laser beam inscription. A standard single-mode fiber (PureBand, G.652.D, Sumitomo Electic) was placed in a hydrogen chamber under high pressure (1500 psi) for one week. Subsequently, the hydrogen-loaded fiber was illuminated by the interferometric fringes of a solid-state laser beam (Impress 213 nm, Xiton photonics) through a phase mask with a pitch of 1082 nm. UV irradiation decomposed the hydrogen, producing a large number of oxidation defects within the fiber core,[28] which resulted in permanent refractive index modulation angled by a few degrees relative to the perpendicular plane of the propagation axis. To excite SPPs on the fiber surface, a 50 nm gold film was applied in two coating runs using a thermal evaporator ($5\times10^{-6}$ Torr, 1 Å/s, Denton).

### 2.3. Real-time biofunctionalization onto the metallic surface

After the gold deposition, the sensor was rinsed several times with Milli-Q water and ethanol separately. For the incubation process, 5 uL of thiolated aptamer (100 μM) was mixed with 5 uL of TECP (10 mM) and incubated at room temperature for 60 mins. The resulting reduced aptamer solution was diluted from 50 μM to 5 μM using 90 uL of PBS, and the 5 μM aptamer solution was then injected into the microfluid tubing using the syringe pump system (ISPLab02, DK Infusetek) for 60 mins. To prevent non-specific adsorption, passivation was performed by immersing the sensor with 2 mM MCH for 45 mins. The entire incubation and passivation process was monitored in real-time by analyzing the spectral data collected from the optical spectrum analyzer (OSA, AQ6370D, Yokogawa).

## 2.4. TFBG-SPR biosensor experimental system

Figure 1A illustrates the experimental setup of the plasmonic TFBG microfluid biosensing system. The setup utilized a superluminescent diode light source with a central wavelength of 1550 nm (MCSLD-1550-20-1, Ming Chuang) and a power accuracy maintained within ±0.05 dB over eight hours. The self-calibration function of the TFBG core mode effectively minimized crosstalk due to temperature and strain variations. SPP is a transverse magnetic (TM) mode and propagates axially along the fiber surface, that is, only the p-polarized light of the evanescent field of the fiber grating cladding mode will be coupled with SPP to excite the SPR. To excite SPR, a polarization controller and a polarizer were employed to generate p-polarized light, enabling efficient coupling with the SPP. The OSA provided a resolution of 0.05 nm, and real-time data collecting was achieved using a laptop via a GPIB interface. Buffer and samples were injected into the sensing system by the two-channel microfluid device, with perfusion and extraction modes, enabling continuous sample changing and buffer cleaning.

## 2.5. DNA tetrahedra assembly

All DNA nanostructures were assembled through an annealing process using the ProFlex PCR System (Applied Biosystems, Foster City, CA, USA). Initially, all strands were activated with TCEP at a ratio of 1:100 for 1 hour. Following activation, equimolar amounts of single-stranded DNA were combined to a final concentration of 10 μM in 1× PBS. The mixture was first denatured at 95°C for 5 minutes, then gradually annealed at a rate of −0.1°C per 8 seconds until it reached 20°C. The assembled nanostructures were stored at 4°C for long-term preservation.

## 2.6. Verification of DNA tetrahedra formation via polyacrylamide gel electrophoresis (PAGE)

To confirm the formation of DNA nanostructures, a 10% PAGE was conducted at 100 V for 1 hour. Each lane was loaded with 100 nM of oligonucleotides. The gel was subsequently stained with Sybr® Gold and visualized using the Gel Doc XR imaging system (Bio-Rad, Hercules, CA, USA).

## 2.7. Optical fiber surface characterization using Atomic Force Microscope (AFM)

The acquisition of AFM images was performed using a Bruker NanoWizard ULTRA Speed 2 AFM, employing the peakforce tapping mode in air environment. SNL-10d AFM tapping mode probes (Bruker Nano, Inc.) with an average force constant of 0.06 N/m and a mean resonance frequency of 18 kHz was used in imaging with scan areas of 1 μm × 1 μm. A scanning resolution of 256 lines with 256 pixels per line with a scan rate of 1 Hz was maintained throughout all image acquisitions. The JPK Data Processing software program was utilized for the meticulous analysis of DNA contour lengths, as previously described.

2.8. Analyze of N protein in Artificial Salivary Matrix

To simulate the detection of N protein in a clinically relevant context, various concentrations of N protein were introduced into an artificial salivary matrix. This synthetic saliva mimics the chemical composition and physical properties of natural saliva, providing a stable environment for in vitro and in vivo studies in pharmacology and bioengineering. Detection was subsequently performed using the TFBG-SPR sensor, which enabled real-time monitoring of N protein levels.

## 3. Results and discussion

3.1. Characterization of aptamer-based TFBG-SPR biosensor

The 18° TFBG fabricated in this study generates a spectrally dense comb of backward propagating central cladding resonance modes, enabling precise measurement of the SRI in the range of 1.33 to 1.35 in aqueous solutions.[29] The central wavelength of 18° TFBG is around 1510 nm, which maximizes the intensity sensitivity since the effective refractive index of the SPR mode is close to aqueous solution. As shown in Figure 1A, the red arrow aligned with the fiber axis represents the propagation of the fundamental mode within the fiber core. When transmitted to the grating region, the TFBG facilitates coupling of the fundamental mode into the cladding, where it propagates backward as cladding modes (illustrated in multiple colors in Figure 1B). Each cladding mode is associated with a distinct effective refractive index with a corresponding unique resonant wavelength. The optical characteristics of gold-coated TFBG can be described by the following two equations:[30]

$$\lambda_{clad,i} = \frac{\left(N_{clad,i}^{eff}+N_{core}^{eff}\right)\Lambda}{cos\theta} \tag{1}$$

$$\beta_{spw} = \frac{\omega}{c}\sqrt{\frac{\varepsilon_s\varepsilon_m}{\varepsilon_s+\varepsilon_m}} \tag{2}$$

In Equation (1), $\lambda_{clad,i}$ represents the resonant wavelength of the high-order cladding modes, while $N_{core}^{eff}$ and $N_{clad,i}^{eff}$ denote the effective refractive indices of the input core mode and excited cladding mode with the order of $i$, respectively, resulting in a comb-shaped spectrum. $\Lambda$ corresponds to the grating period and $\theta$ is the tilt angle. In Equation (2), $\omega$, $\beta$, and $c$ represent the angular frequency of light, axial component of the propagation constant, and the speed of light in vacuum, respectively. Additionally, $\varepsilon_s$ and $\varepsilon_m$ are the complex relative permittivities of the surrounding medium and the metal film. The SPR mode can be theoretically predicted using Equation (2). As described, the evanescent wave of the cladding modes and the SPP waves satisfy the phase-matching condition $\beta_{spw} = \beta_{clad,i}$. Based on this mechanism, the SPR signal is tracked to monitor and quantify the SRI.

When the gold-coated TFBG is immersed in aqueous solutions, the propagation constants of the excited SPPs are determined by the refractive index of analyte in near infrared band. As the dielectric samples attach to the surface of metallic TFBG, such as the N protein, DNA tetrahedra (DNT), aptamer, then the complex relative permittivity of the metal $\varepsilon_m$ will increase, and the plasmon resonance wavelength will red shift accordingly.

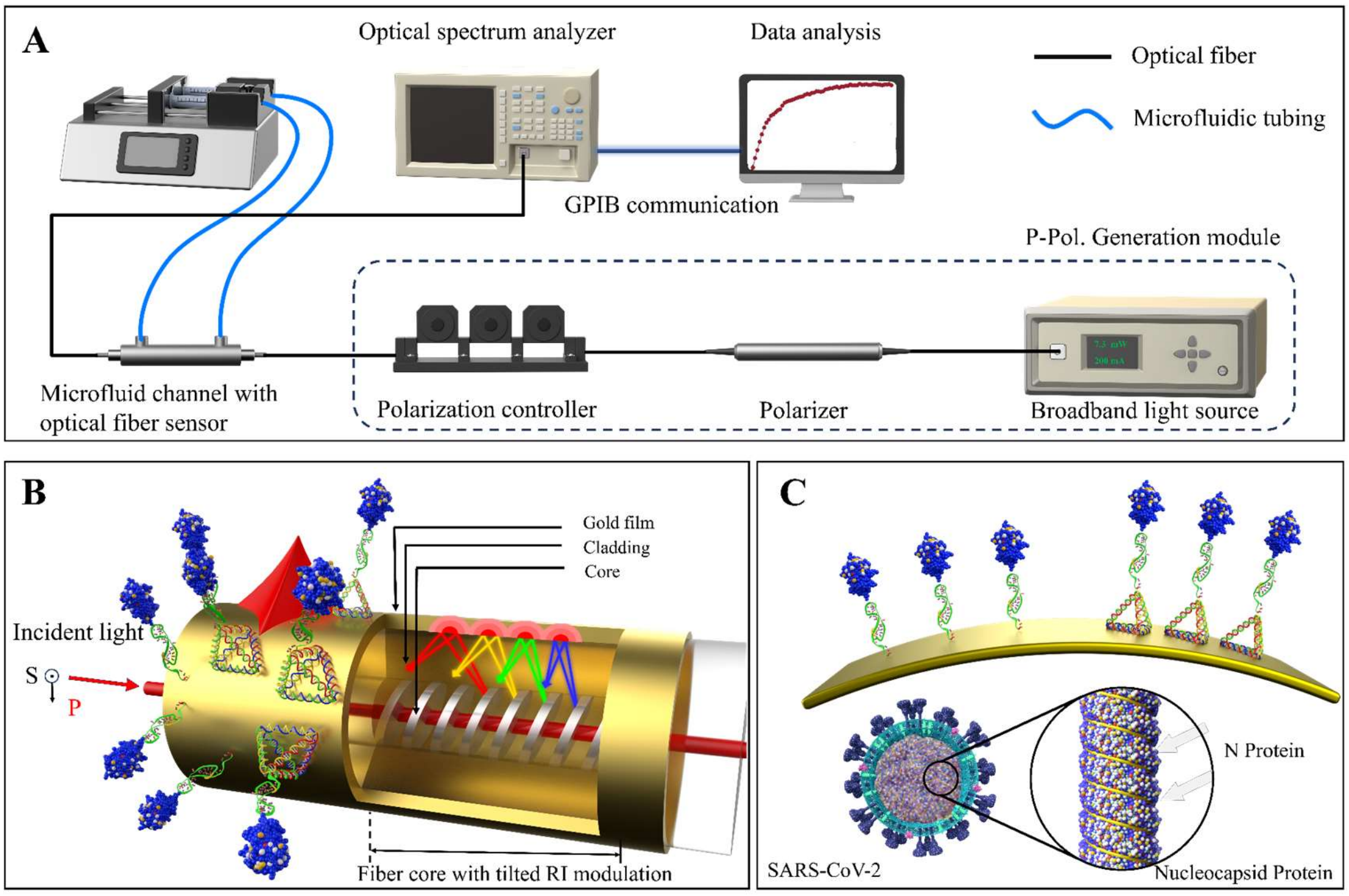

Figure 1. Schematic framework of the TFBG-SPR sensing system. (A) Design of the plasmonic TFBG microfluidic biosensing system, including p polarization light generation, data processing and microfluid system. (B) Schematic of the TFBG-SPR sensing mechanism, showing the excited cladding modes (rainbow arrows) coupled into a decayed SPR field. RI: Refractive Index. (C) Interface of the gold-coated fiber surface with single aptamer/aptamer-DNA tetrahedra for N protein capture.

## 3.2. Working principle of the aptamer-based TFBG-SPR biosensing system

Two strategies of the receptor we employed are shown in Figure 1C. N-protein specific aptamer (named Apt48) was used to develop the aptamer-based TFBG-SPR biosensor,[31] the sequence attached in Table S1. The single stranded aptamer and aptamer-DNA tetrahedra were separately incubated onto the golden surface. Two surfaces were developed – one with aptamer alone and one with aptamer-tetrahedron.

With the operation in polarization controlling module (Polarizer and Polarization controller), we manipulated the broadband beam from SLD into the p polarized state where radially polarized cladding modes are excited. After the injected solutions cover the optical fiber sensor, as shown in the spectrum Figure 2 collected by optical spectrum analyzer (OSA), the SPR area was excited to where the effective refractive index of cladding mode is around 1.322, which is approximately 1510 nm using the calculation of equation 1. With the help of algorithm that we developed for OSA communication and response intensity trace in matlab, the biosensing system plotted every injection of solution by analyzing the spectral data real time transformed into the computer. The most sensitive sensing pair towards SRI change were selected from the SPR area among the cladding modes, meanwhile, the core mode was collected to monitor the stability of the sensing system (Figure S1 A-F).

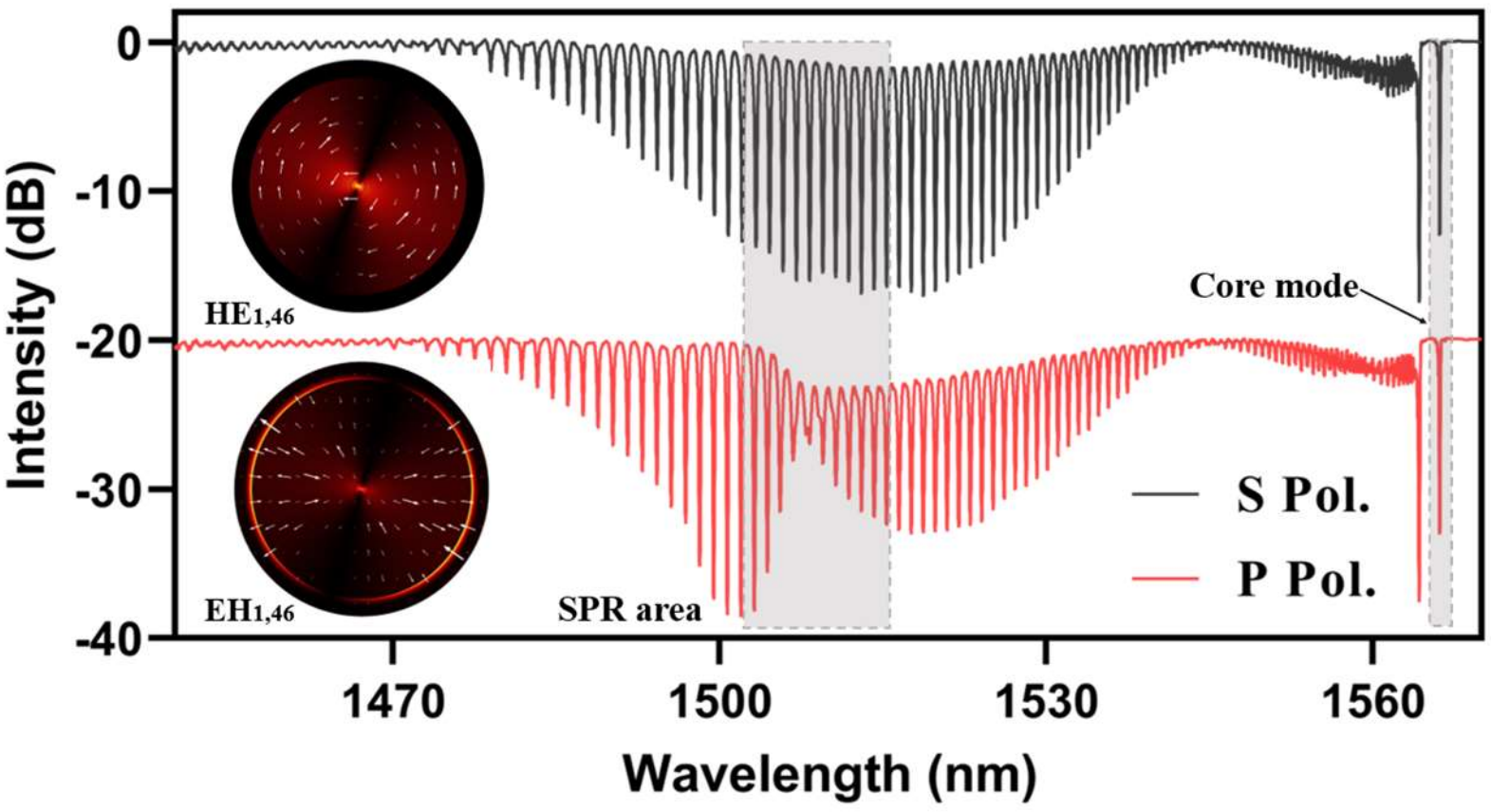

Figure 2. Spectrum of TFBG-SPR in p polarization and s polarization. The left rectangle region in shorter wavelength indicates the specified cladding modes near the phase-matching of the solution based SPP mode, while the right rectangle region labels the core mode which provides the function of monitoring the stability of the sensing system. The hybrid modes $HE_{1,46}$ and $EH_{1,46}$ are inserted to illustrated the strong SPR only occurs on the surface of the cladding modes with the electric field azimuthally radiated.

### 3.3. DNA nanostructure assembly and immobilization

In this study, to compare the performance of DNA nanostructure and single aptamers on the fiber surface, we designed a DNA tetrahedron (DNT) with an edge length of 17 bases using Tiamat.[32] Three of the strands were thiol-labeled to enable conjugation to the gold surface, and the fourth strand contained an extended Apt48 sequence (DNT-Apt48), with sequence information provided in Supporting Information Table 1. The aptamer- tetrahedron is illustrated in Figure 3A. The assembly of the nanostructures was observed through PAGE (Figure 3B). The assembly of single-stranded DNA met expectations, with the size increasing as the number of strands increased. The successful immobilization of DNT-Apt48 on the fiber surface was also clearly demonstrated by comparing the AFM image to bare fiber controls, as shown in Figure 3C and D. The step-by-step process for the preparation of the sensing bio-layer is shown in Figure 3E.

### 3.4. Feasibility verification of aptamer-based TFBG-SPR biosensing system

In the TFBG-SPR biosensing system, changes in the equilibrium binding of biomolecules lead to variations in surface refractive index, thereby quantifying their concentration. Specifically, by real-time monitoring of the differential amplitude of sensing pairs selected from the SPR region,[33] as shown in Figure 3F, we can detect biofunctionalization and immobilization reactions at specific concentrations in real-time. In detail, the thiol-labeled DNT-Apt48 is first immobilized onto a gold-coated optical fiber via Au–S bonds to achieve surface functionalization. Subsequently, the fiber is incubated with an MCH (6-mercapto-1-hexanol) solution to block nonspecific binding sites and passivate the surface, thereby enhancing the signal-to-noise ratio. Finally, the aptamer captures the target efficiently through its high binding affinity to the N protein.

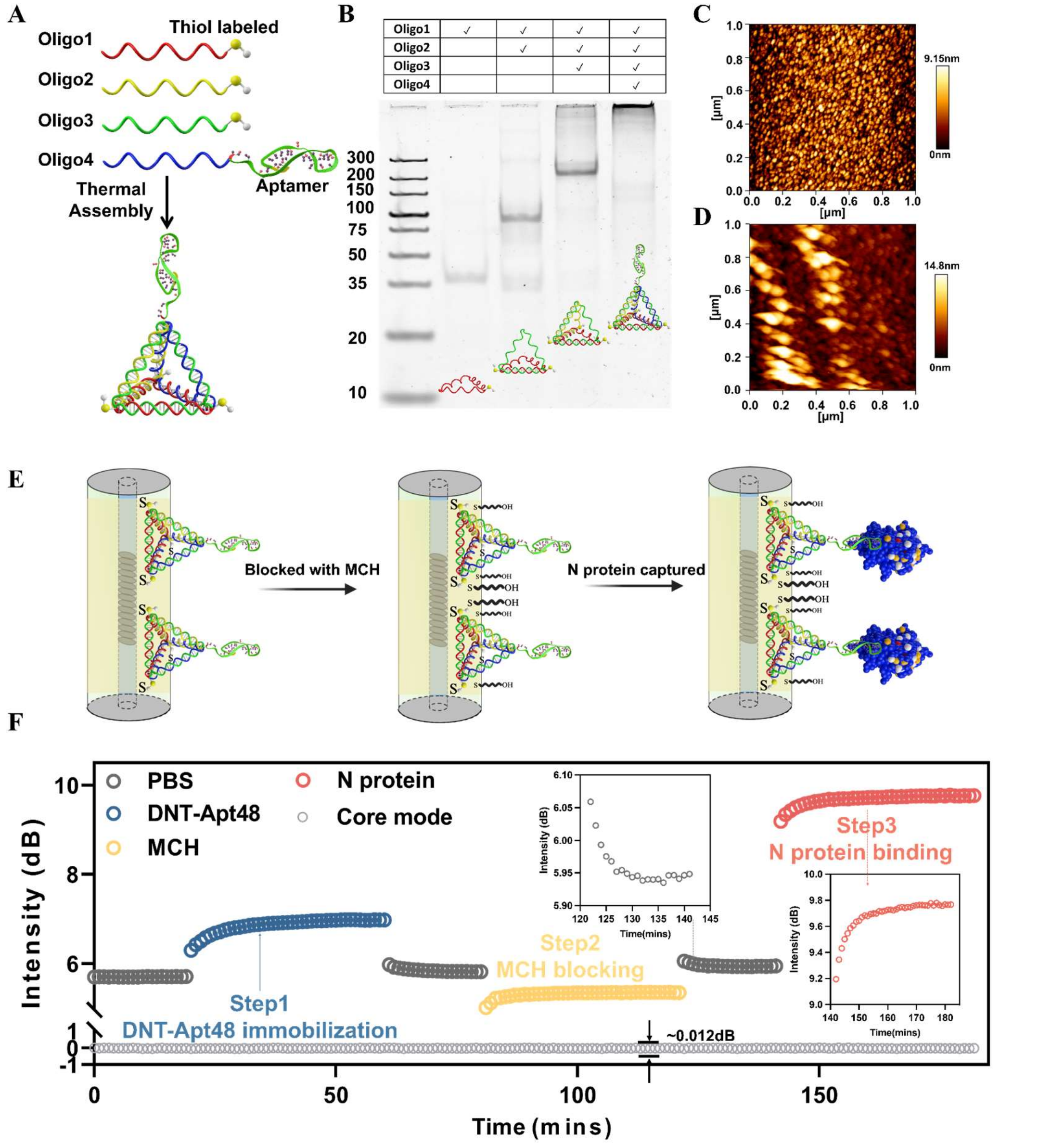

Figure 3. Thermal assembly of the tetrahedron-aptamer nanostructure and surface biofunctionalization of the optical biosensor. (A) Schematic illustration of the DNT-Apt48 assembly process; (B) PAGE gel showing the stepwise assembly of DNT-Apt48; (C) and (D) AFM images comparing the immobilization of DNT-Apt48 on the optical fiber with the bare gold surface. (E) Schematic illustration of immobilizing DNT-Apt48 and the MCH blocking agent on the fiber surface for N protein detection; (F) Real-time signal response of the whole biofunctionalization, the core mode is plotted with light gray line, the mean square error 0.012 dB indicates that the consistency of our SRI sensing results.

### 3.5. Specificity Evaluation and Conditions Optimization

Additionally, to validate the specificity of the aptamer-based TFBG-SPR sensor for detection, we conducted the following experiments. Different from the above real-time detection in Figure 3, for clear comparison, we normalized corresponding signals in Figure 4 with the intensity of first signal and recorded as relative intensity. As shown in Figure 4A, several control groups were set up. First, a control group without Apt48 was tested, and no signal was observed. After immobilizing thiol labeled Apt48 on the fiber, it was incubated with the target N protein or the control S protein, showing a binding signal only for the N protein. Additionally, no signal change was observed when a random sequence control was immobilized on the fiber and incubated with the N protein. All these experiments collectively demonstrate successful thiol conjugation of Apt48 onto the gold-coated optical fiber and the high binding specificity against the N protein, confirming the aptamer-based TFBG-SPR sensor's potential for reliable and specific detection of target biomolecules.

Based on the real time monitoring property of our sensing system, we measured the change in binding signal from incubating 2.5 μM N protein onto 10 μM Apt48 and DNT-Apt48 treated golden surface over time. As shown in Figure 4B, plateau signals were reached at around 25 minutes and the steady-state error was within 0.022 dB, then we collected those signal intensities and fitted them with logistic regression. The aptamer coating concentration were optimized by soaking 5 freshly biofunctionalized plasmonic TFBG with sequential receptors (Apt48 and DNT-Apt48 in 1 μM, 2 μM, 5 μM, 7.5 μM and 10 μM) treated with 0.4 μM N protein. The corresponding signals were normalized with the intensity of first signal for clear comparison. From Figure 4C, the N protein binding signals were nearly two times amplified. For supplying sufficient binding sites for capturing N protein, we chose the concentration of 10 μM for our following receptor solutions.

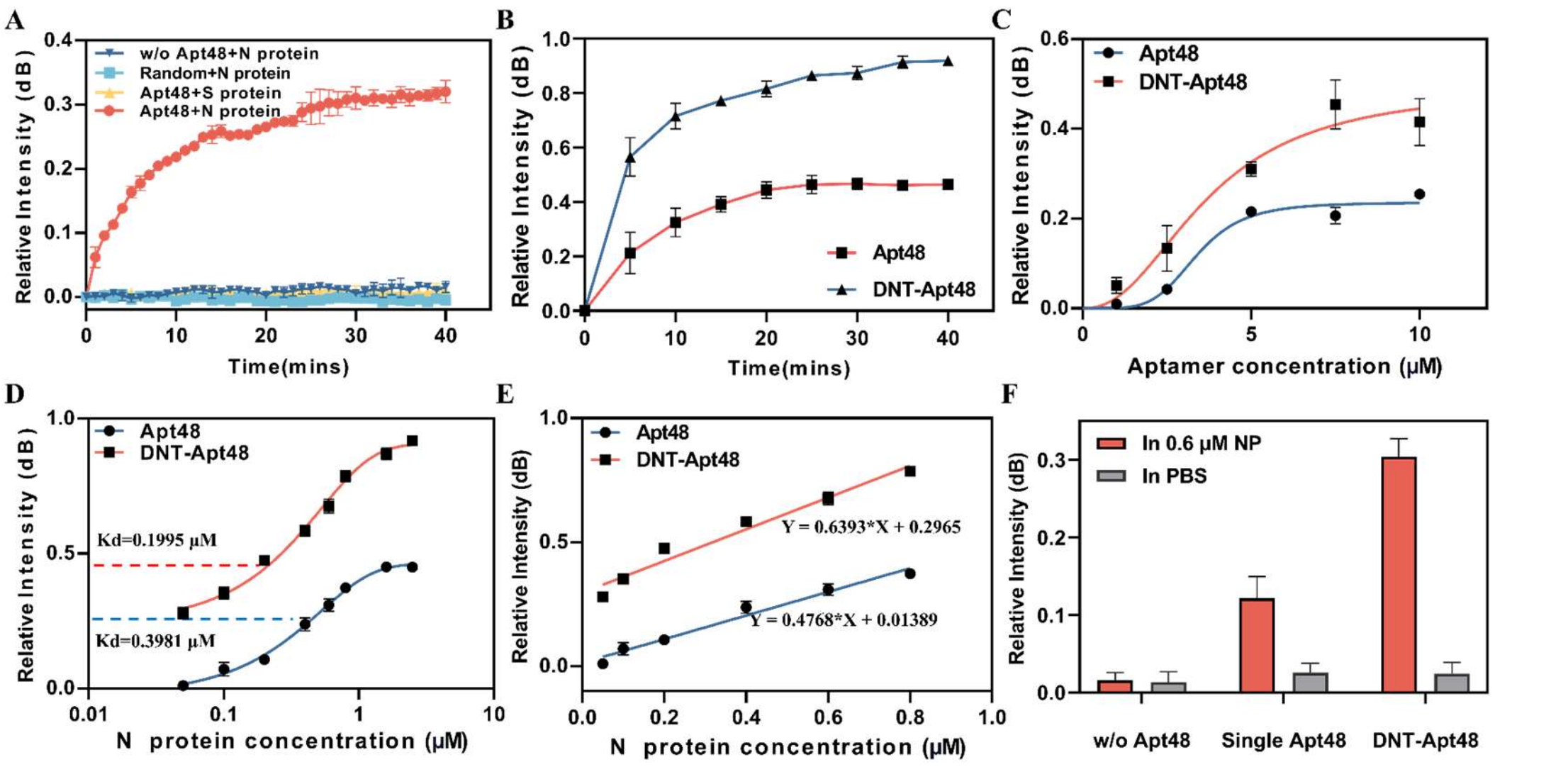

Figure 4. Specificity of N protein sensing, signal amplification with the DNA tetrahedron surface modification, and detection in salivary matrix (A) Specificity verification of Apt48 binding to 0.6 μM N protein. (B) Real-time optimization of binding time (2.5 μM N protein with 10 μM Apt48). (C) Concentration optimization of single Apt48 and DNT-Apt48. (D) Sensitivity comparison of single Apt48 and DNT-Apt48 for capturing N protein on the fiber. (E) Limit of detection (LOD) calculation for single Apt48 and DNT-Apt48, where LOD = 3 × σ / slope (σ: standard deviation of the blank control; Apt48: 46 nM, DNT: 34 nM). (F) Detection of 0.6 μM N protein in artificial saliva (N = 3).

### 3.6. Detection of N protein by biofunctionalized optical fiber

By preparing seven sets of N protein solutions with concentration gradients ranging from 0.05 to 2.5 μM, we observed in Figure 4D that the incorporation of DNT enhances the sensitivity of the biosensor when combined with Apt48. Specifically, the dissociation constant (Kd) of Apt48 alone was determined to be 0.3981 μM, whereas the Kd of the DNT-Apt48 complex was reduced to 0.1996 μM. To further evaluate the practical detection performance of the biosensor, we selected concentrations close to the Kd for response calibration. Six groups of N protein solutions were prepared around the Kd value (~0.4 μM), and their corresponding signal responses were linearly fitted, as shown in Figure 4E. For practical applications, we localized the linear fitting region around the Kd, where approximately 50% of the binding sites are occupied, rather than extending it to extremely low concentrations that typically exhibit a steeper response slope but suffer from a poor signal-to-noise ratio (Figure S2). This approach ensures more reliable and reproducible quantification within the sensor’s effective dynamic range, rather than focusing solely on minimizing the limit of detection (LOD).We can see that the signal intensity is amplified by nearly two times, and at a concentration of 0.05 µM for N protein, the sensor incubated with Apt48 alone is only 0.01 dB, which is already the limit of resolution in our system, while the sensor incubated with DNT still has a signal response of about 0.28 dB.

To demonstrate whether the aptamer-based TFBG-SPR sensor retains its detection capability in complex environments and to compare the performance of single Apt48 and DNT-Apt48, we introduced N protein into artificial saliva to simulate human bodily fluids. As shown in Figure 4F, we added 0.6 µM N protein to the artificial saliva and incubated it with fibers functionalized with 10 µM single Apt48 and DNT-Apt48, respectively. The results indicated that the TFBG-SPR sensor could still detect signals even when the target was in a complex sample. Additionally, compared to the single Apt48, the DNT-Apt48 exhibited superior target capture performance, reaching a plateau within 25 minutes, with signal intensity approximately twice that of the single aptamer.

### 3.7. Regeneration of optical fiber surface

Subsequently, the surface reproducibility of the sensor was evaluated. A 1% w/v SDS solution was used to elute the proteins bound to the aptamer without disrupting the aptamer's immobilization. After rinsing with PBS, the aptamer regained its secondary structure and was able to rebind with the N protein. As shown in Figure 5 and Figure S3, two sets of regeneration experiments were performed, each consisting of three cycles. In the first set as shown in Figure 5, the concentration of N protein was gradually increased over three cycles, and significant signal changes were still observed. In the second set as shown in Figure S3, the concentration of N protein was kept constant, and three regeneration cycles were carried out, demonstrating a regeneration efficiency of up to 97%.

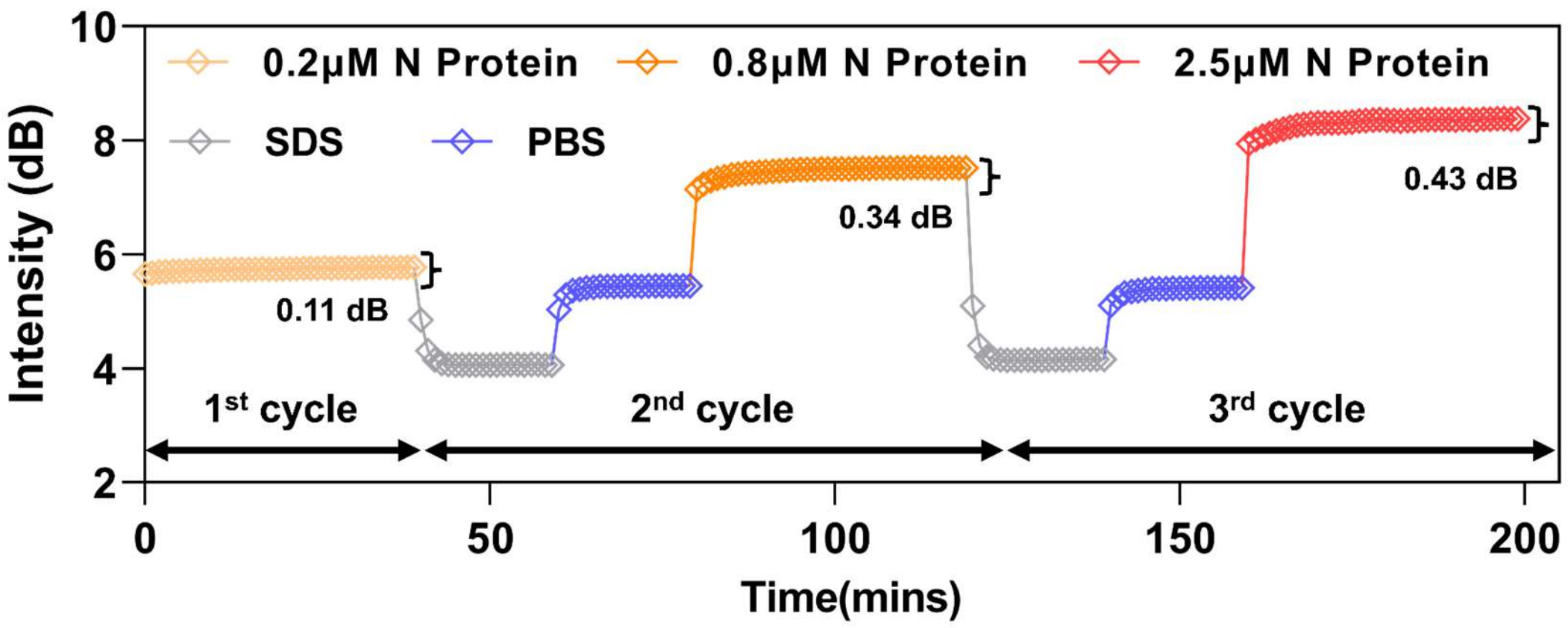

Figure 5. Surface reproducibility of the biosensor, renewability with the protein solubilization. The sensor was reused over three detection cycles with increasing concentrations of N protein (0.2 µM, 0.8 µM, and 2.5 µM), each followed by regeneration using 1% w/v SDS. Distinct signal responses were retained across all cycles.

## 4. Conclusion

In conclusion, we designed and developed an aptamer-based TFBG-SPR sensor for the detection and identification of disease biomarkers. The binding of the target induces changes in the surface refractive index, generating detectable signals. In this study, we selected Apt48, which specifically recognizes the N protein of SARS-CoV-2 and immobilized it on the gold-coated TFBG surface via Au-S bonding. To verify the feasibility and specificity of the Apt48-N protein complex, we conducted control experiments using a random sequence aptamer and a non-specific target (S protein) as substitutes for the receptor or target. The experimental results confirmed the specific binding capability of Apt48 to the N protein, demonstrating its potential for further applications Furthermore, based on our previous studies, DNA nanostructures can effectively display aptamers. We designed DNT-Apt48 with 17-nt edges and compared their performance with single Apt48 immobilized on the gold-coated optical fiber. The experimental results showed that DNT-Apt48 exhibited faster signal responses and a 2~2.5-fold increase in signal intensity compared to single Apt48. Additionally, by varying the N protein concentration, we calculated and compared the Kd value of single Apt48 and DNT-Apt48, which were 0.398 μM and 0.199 μM, respectively. Overall, the aptamer-TFBG-SPR sensor successfully detected the N protein, and the use of DNA nanostructures to display aptamers enhanced the signal intensity, providing a novel strategy for further applications of optical fiber sensors. To evaluate the performance of the TFBG-SPR sensor in complex environments, we tested N protein in artificial saliva. The results demonstrated that the sensor retains excellent detection capabilities in complex environments, with DNT-Apt48 exhibiting stronger signal responses, further confirming its superiority and practicality.

Additionally, the regeneration capability of the optical fiber surface biosensor was demonstrated, allowing the sensor to be reused multiple times without significant performance degradation. By soaking and rinsing the fiber surface with 1% w/v SDS solution, the binding between the aptamer and the target was disrupted. The sensor was then reintroduced to the target for subsequent detection, with the regenerated sensor retaining over 97% of its original signal strength. This is attributed to the high stability and structural switching properties of the aptamer. Such regeneration not only enhances the cost-effectiveness of the sensor but also highlights its practical value and long-term potential in biomolecular detection applications. Surface modification of nanoplasmonic biosensors with aptamers for molecular recognition coupled to DNA nanostructure conjugation layers could hold promise in a wide range of biosensing applications.

**Supporting Information**

Supporting information is available from the Wiley Online Library or from the author

**Acknowledgements**

This work was supported in part by the Research Grants Council (RGC) of Hong Kong (15209321 B-Q85G), the Hong Kong Polytechnic University (1-CD5D). This work was also supported by RGC GRF grants 17125920, 17125221, 17127124 and TBRS grant T12-201/19-R, as well as ITC ITF grant SST/118/20GP to JAT.

**Conflict of Interest**

The authors declare no conflict of interest.

**Data Availability Statement**

The data that support the findings of this study are available from the corresponding author upon reasonable request.

**References**

[1] S. Poolsup, E. Zaripov, N. Hüttmann, Z. Minic, P. V. Artyushenko, I. A. Shchugoreva, F. N. Tomilin, A. S. Kichkailo, M. V. Berezovski, *Molecular Therapy-Nucleic Acids* **2023**, 31, 731.

[2] J. E. Sanchez, S. A. Jaramillo, E. Settles, J. J. Velazquez Salazar, A. Lehr, J. Gonzalez, C. Rodriguez Aranda, H. R. Navarro-Contreras, M. O. Raniere, M. Harvey, D. M. Wagner, A. Koppisch, R. Kellar, P. Keim, M. Jose Yacaman, *Rsc Advances* **2021**, 11, 25788.

[3] W. Wang, S. Srivastava, A. Garg, C. Xiao, S. Hawks, J. Pan, N. Duggal, G. Isaacman-VanWertz, W. Zhou, L. C. Marr, P. J. Vikesland, *Environ. Sci. Technol.* **2024**, 58, 4926.

[4] A. A. Hariri, A. P. Cartwright, C. Dory, Y. Gidi, S. Yee, I. A. Thompson, K. X. Fu, K. Yang, D. Wu, N. Maganzini, *Adv. Mater.* **2024**, 36, 2304410.

[5] Y. Wang, X. Long, X. Ding, S. Fan, J. Cai, B. Yang, X. Zhang, R. Luo, L. Yang, T. Ruan, J. Ren, C. Jing, Y. Zheng, X. Hao, D. Chen, *European Journal of Medicinal Chemistry* **2022**, 227, 113966.

[6] T.-a. Yano, T. Kajisa, M. Ono, Y. Miyasaka, Y. Hasegawa, A. Saito, K. Otsuka, A. Sakane, T. Sasaki, K. Yasutomo, R. Hamajima, Y. Kanai, T. Kobayashi, Y. Matsuura, M. Itonaga, T. Yasui, *Scientific Reports* **2022**, 12, 1060.

[7] Y. M. Aloraij, G. A. Suaifan, A. Shibl, K. Al-Kattan, M. M. Zourob, *ACS omega* **2023**, 8, 32877.

[8] J. Homola, *Analytical bioanalytical chemistry* **2003**, 377, 528.

[9] J. Cao, M. H. Tu, T. Sun, K. T. Grattan, *Sensors Actuators B: Chemical* **2013**, 181, 611.

[10] R. Slavık, J. Homola, E. Brynda, *Biosens. Bioelectron.* **2002**, 17, 591.

[11] H.-Y. Lin, C.-H. Huang, G.-L. Cheng, N.-K. Chen, H.-C. Chui, *Opt. Express* **2012**, 20, 21693.

[12] V. Sai, T. Kundu, S. Mukherji, *Biosens. Bioelectron.* **2009**, 24, 2804.

[13] J. Albert, L. Y. Shao, C. J. L. Caucheteur, P. Reviews, **2013**, 7, 83.

[14] X. Chen, Y. Nan, X. Ma, H. Liu, W. Liu, L. Shi, T. Guo, *J. Lightwave Technol.* **2019**, 37, 2792.

[15] L. Liu, X. Zhang, Q. Zhu, K. Li, Y. Lu, X. Zhou, T. Guo, *Light Sci. Appl.* **2021**, 10, 181.

[16] C. Ji, J. Wei, L. Zhang, X. Hou, J. Tan, Q. Yuan, W. Tan, *Chem. Rev.* **2023**, 123, 12471.

[17] Z. Chen, H. Luo, A. Gubu, S. Yu, H. Zhang, H. Dai, Y. Zhang, B. Zhang, Y. Ma, A. Lu, G. Zhang, *Frontiers in Cell and Developmental Biology* **2023**, 11, 1091809.

[18] M. R. Dunn, R. M. Jimenez, J. C. Chaput, *Nature Reviews Chemistry* **2017**, 1, 0076.

[19] A. D. Ellington, J. W. Szostak, *Nature* **1990**, 346, 818.

[20] C. Tuerk, L. Gold, *Science* **1990**, 249, 505.

[21] N. C. Seeman, *Journal of Theoretical Biology* **1982**, 99, 237.

[22] R. P. Goodman, I. A. T. Schaap, C. F. Tardin, C. M. Erben, R. M. Berry, C. F. Schmidt, A. J. Turberfield, *Science* **2005**, 310, 1661.

[23] S. C.-C. Shiu, L. A. Fraser, Y. Ding, J. A. Tanner, *Molecules* **2018**, 23, 1695.

[24] F. Wu, S. Zhao, B. Yu, Y.-M. Chen, W. Wang, Z.-G. Song, Y. Hu, Z.-W. Tao, J.-H. Tian, Y.-Y. Pei, M.-L. Yuan, Y.-L. Zhang, F.-H. Dai, Y. Liu, Q.-M. Wang, J.-J. Zheng, L. Xu, E. C. Holmes,

Y.-Z. Zhang, *Nature* **2020**, 579, 265.

[25] A. W. D. Edridge, J. Kaczorowska, A. C. R. Hoste, M. Bakker, M. Klein, K. Loens, M. F. Jebbink, A. Matser, C. M. Kinsella, P. Rueda, M. Ieven, H. Goossens, M. Prins, P. Sastre, M. Deijs, L. van der Hoek, *Nat. Med.* **2020**, 26, 1691.

[26] A. Cetinkaya, S. I. Kaya, S. A. Ozkan, *Crit. Rev. Anal. Chem.* **2024**, 54, 2517.

[27] H. M. Rando, C. Brueffer, R. Lordan, A. A. Dattoli, D. Manheim, J. G. Meyer, A. I. Mundo, D. Perrin, D. Mai, N. J. A. Wellhausen, *ArXiv [Preprint]* **2022**.

[28] P. Karlitschek, G. Hillrichs, K. F. Klein, *Opt. Commun.* **1998**, 155, 386.

[29] T. Guo, F. Liu, B.-O. Guan, J. Albert, *Opt. Laser Technol.* **2016**, 78, 19.

[30] R. Wang, H. Zhang, Q. Liu, F. Liu, X. Han, X. Liu, K. Li, G. Xiao, J. Albert, X. Lu, T. Guo, *Nature Communications* **2022**, 13.

[31] L. Zhang, X. Fang, X. Liu, H. Ou, H. Zhang, J. Wang, Q. Li, H. Cheng, W. Zhang, Z. Luo, *Chem. Commun.* **2020**, 56, 10235.

[32] S. Williams, K. Lund, C. Lin, P. Wonka, S. Lindsay, H. Yan, presented at DNA Computing: 14th International Meeting on DNA Computing, DNA 14, Prague, Czech Republic, June 2-9, 2008. Revised Selected Papers 14 **2009**.

[33] X. Chen, P. Xu, W. Lin, J. Jiang, H. Qu, X. Hu, J. Sun, Y. Cui, *Biomedical Optics Express* **2022**, 13, 2117.